\documentclass[twocolumn]{aastex701}

\usepackage{fontawesome}
\usepackage{hyperref}

\newcommand{\githubicon}{{\color{black}\faGithub}}

\begin{document}

\title{Precovery Observations of 3I/ATLAS from TESS Suggests Possible Distant Activity}

\author[orcid=0000-0000-0000-0001,sname='Feinstein']{Adina~D.~Feinstein}
\altaffiliation{NASA Sagan Fellow}
\affiliation{Department of Physics and Astronomy, Michigan State University, East Lansing, MI 48824, USA}
\email[show]{adina@msu.edu}  

\author[0000-0003-2152-6987]{John W.\ Noonan}
\affiliation{Department of Physics, Auburn University, Edmund C.\ Leach Science Center, Auburn, 36849, AL, USA}
\email{noonan@auburn.edu}  

\author[0000-0002-0726-6480]{Darryl Z. Seligman}
\altaffiliation{NSF Astronomy and Astrophysics Postdoctoral Fellow}
\affiliation{Department of Physics and Astronomy, Michigan State University, East Lansing, MI 48824, USA}
\email{dzs@msu.edu}

\begin{abstract}

3I/ATLAS is the third macroscopic interstellar object detected traversing the Solar System. Since its initial discovery on UT 01 July 2025, hundreds of hours on a range of observational facilities have been dedicated to measure the physical properties of this object. These observations have provided astrometry to refine the orbital solution, photometry to measure the color, a rotation period and secular light curve, and spectroscopy to characterize the composition of the coma. Here, we report precovery photometry of 3I/ATLAS as observed with NASA's \textit{Transiting Exoplanet Survey Satellite} (\textit{TESS}). 3I/ATLAS was observed nearly continuously by \textit{TESS} from UT 07 May 2025 to 02 June 2025. We use the shift-stack method to create  deep stack images to recover the object. These composite images reveal that 3I/ATLAS has an average \textit{TESS} magnitude of $T_\textrm{mag} = 20.83 \pm 0.05, 19.28 \pm 0.05$ and an absolute visual magnitude of $H_V = 13.72 \pm 0.35; 12.52 \pm 0.35$, the latter being consistent with magnitudes reported in July 2025. When coupled with recent HST images deriving a nucleus size of R$<$2.8 km (H$>$15.4), our measurements suggest that 3I/ATLAS may have been active out at $\sim 6$~au. Additionally, we extract a $\sim 20$~day light curve and find no statistically significant evidence of a nucleus rotation period. Nevertheless, the data presented here are some of the earliest precovery images of 3I/ATLAS and may be used in conjunction with future observations to constrain the properties of our third interstellar interloper.

\end{abstract}


\section{Introduction} 

Interstellar objects (ISOs) offer unique insights into the formation conditions of rocks and ices  in regions of our  galaxy beyond the  Solar System. The recent discovery of the third ISO, 3I/ATLAS, was reported on UT 01 July 2025 \citep{Denneau2025} by the ATLAS survey \citep{Tonry2018a,Tonry2018b}. This discovery increases the current census of discovered macroscopic ISOs by 50\%. Since its discovery, there has been an international rush to observe 3I/ATLAS before it is no longer visible to observers in February 2026. In particular, early telegrams reported activity in images from the Nordic-Optical Telescope \citep{Jewitt2025} and the Two-meter Twin Telescope \citep{Alarcon2025}. Preliminary reconnaissance observations revealed a reddened reflectance spectrum similar to D-type asteroids, no clear photometric variability and  faint cometary activity \citep{Kareta2025,Marcos2025,Opitom2025,seligman25,Yang2025}.

Two macroscopic scale ISOs had been discovered before 3I/ATLAS: 1I/`Oumuamua in 2017 \citep{Williams17} and 2I/Borisov in 2019 \citep{borisov_2I_cbet}. The light curve of 1I/`Oumuamua was markedly different from the 3I/ATLAS because it exhibited drastic brightness variations \citep{Drahus2017,Fraser2017,Knight2017,Belton2018}. This was presumably due to an elongated $6:6:1$ oblate spheroid geometry \citep{Mashchenko2019,Taylor2023} and a lack of cometary activity \citep{Jewitt2017,Meech2017,Ye2017}. A significant non-detection of 1I/`Oumuamua with the \textit{Spitzer} space telescope provided upper limits on the production of carbon-based species \citep{Trilling2018}. However, 1I/`Oumuamua had significant radial nongravitational acceleration in its trajectory \citep{Micheli2018}. 1I/`Oumuamua had a reddened color like 3I/ATLAS \citep{Bannister2017,Fitzsimmons2017,Masiero2017} and a low velocity at infinity indicating a young kinematic age \citep{Gaidos2017a,Mamajek2017}. On the contrary, 2I/Borisov displayed distinct cometary activity \citep{Jewitt2019b,Fitzsimmons:2019,Kareta:2019,Opitom:2019-borisov, Bannister2020,Bodewits2020,Cremonese2020,Cordiner2020,Guzik:2020,Hui2020,Kim2020,lin2020,McKay2020,Xing2020,Aravind2021, Bagnulo2021,yang2021,Deam2025}. For reviews on ISOs discovered passing through the solar system prior to 3I/ATLAS, we refer the reader to \citet{MoroMartin2022,Jewitt2023ARAA,Seligman2023,Fitzsimmons2024}.

Precovery observations of small bodies such as comets, asteroids, and ISOs can provide constraints on critical physical properties such as orbit, size, composition, and activity. For example, \citet{Hui2019} searched for precovery detections of 1I/`Oumuamua in Solar and Heliospheric Observatory (SOHO) and Solar TErrestrial RElations Observatory (STEREO) images. Their search provided nondetections, but these nondetections provided critical upper limits on the production rate of dust and water at  low heliocentric distances --- during a regime of 1I/`Oumuamua's trajectory which was entirely unconstrained from extant observations. \citet{Ye2020} reported pre-discovery observations of 2I/Borisov in survey data from the Zwicky Transient Facility (ZTF), as well as a comprehensive search of Catalina Sky Survey and Pan-STARRS data for the object. The combination of precovery detections and nondetections provided valuable upper limits on the size of the nucleus and information regarding the volatility of ices driving activity. Thus far for 3I/ATLAS, \cite{seligman25} reported ZTF observations dating back to UT 2025 May 22 and \citet{Chandler2025} reported observations of using the NSF-DOE Vera C. Rubin Observatory Science Verification (SC) images dating back to UT 2025 June 21. The Rubin Observatory SC observations also revealed faint activity. 

All four aforementioned studies highlight the strengths of all-sky surveys in searching for precovery observations. Although not its original science purpose, NASA's \textit{Transiting Exoplanet Survey Satellite} \citep[\textit{TESS};][]{ricker15} has also been able to recover and discover small bodies in the solar system  \citep{pal20, szabo22, kiss25} as well as characterize cometary activity \citep{farnham2019_46ptess,farnham2021_un271tess}. \textit{TESS} has a large on-sky coverage due to the orientation of four wide-field cameras, where each camera has a field-of-view of $24^\circ \times 24^\circ$, amounting to a total FOV of $96^\circ \times 24^\circ$. These images, known as the \textit{TESS} Full-Frame Images (FFIs), are made available to the community. \textit{TESS} observes a single field for $\sim27$~days, providing a long photometric baseline for variability studies. While limited in its resolution for morphology analysis of cometary comae due to its large pixel scale, \textit{TESS} is excellent for long-term photometric monitoring and analysis.

During the primary mission, \textit{TESS} exclusively observed the northern and southern ecliptic hemispheres. However, in its extended missions, \textit{TESS} was reoriented to observe the ecliptic plane for 10 nonconsecutive sectors. One of these sectors --- both fortuitously and serendipitously --- happened to observe the exact region where 3I/ATLAS was located. Here, we report the precovery detection of 3I/ATLAS in the \textit{TESS} Full-Frame Images (FFIs). This manuscript is organized as follows. In Section~\ref{sec:obs}, we present the \textit{TESS} observations and discuss our processing of the FFIs. We present the results of our search in Section~\ref{sec:discuss}. We conclude in Section~\ref{sec:conclusion}. The \githubicon\ icon affiliated with all figures links to the Python script and data used to generate that figure.

\section{TESS Observations}\label{sec:obs}

\textit{TESS} re-observed part of the ecliptic plane during Cycle 7 of its second extended mission. Upon discovery of 3I/ATLAS, we used \texttt{tess-point} \citep{burke20} in conjunction with 3I/ATLAS' best-fit orbital solution from the JPL Horizons Small Bodies Database\footnote{\url{https://ssd.jpl.nasa.gov/}} to determine if the object was in the \textit{TESS} FOV. We found that 3I/ATLAS was observed during Sector 92, which occurred from UT 07 May through 03 June 2025. 

\begin{figure}[hbt!]
\includegraphics[width=1.\linewidth]{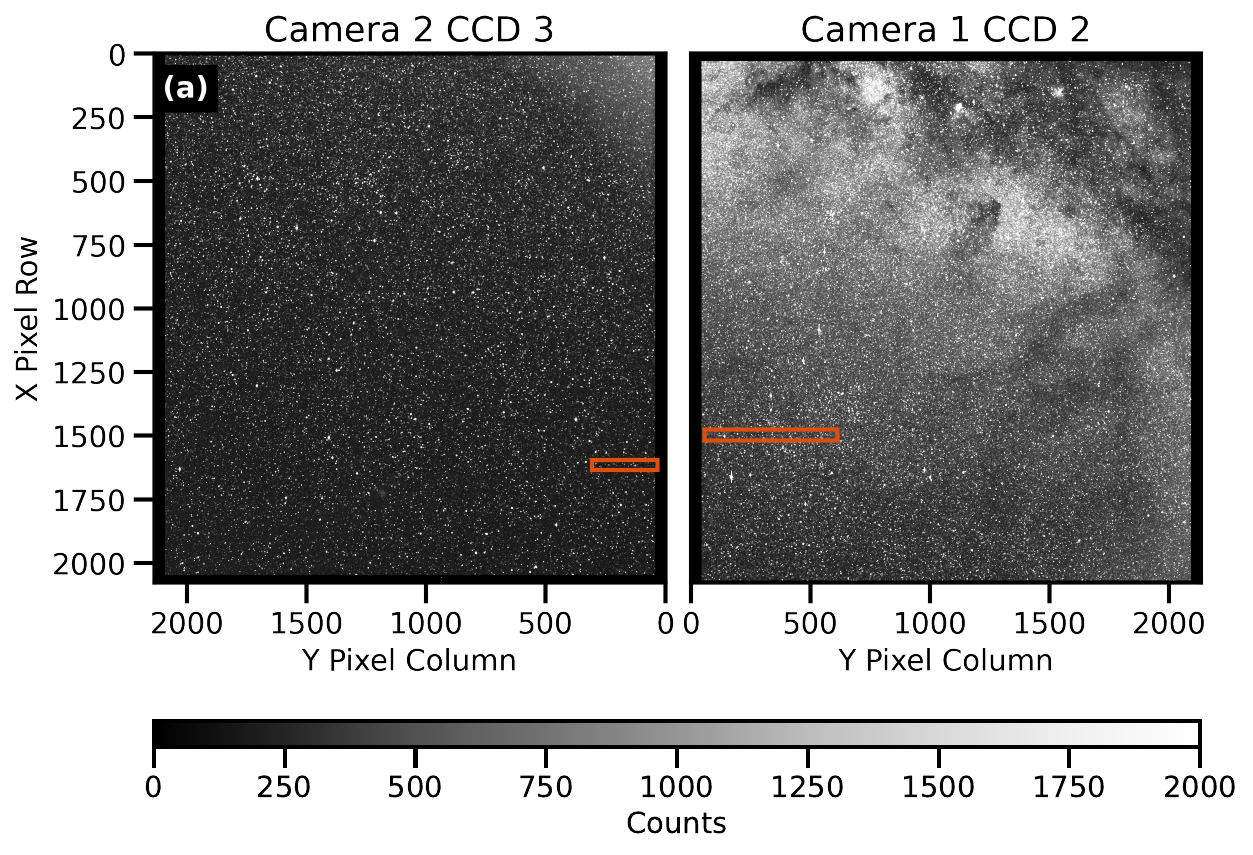}
\includegraphics[width=1.\linewidth]{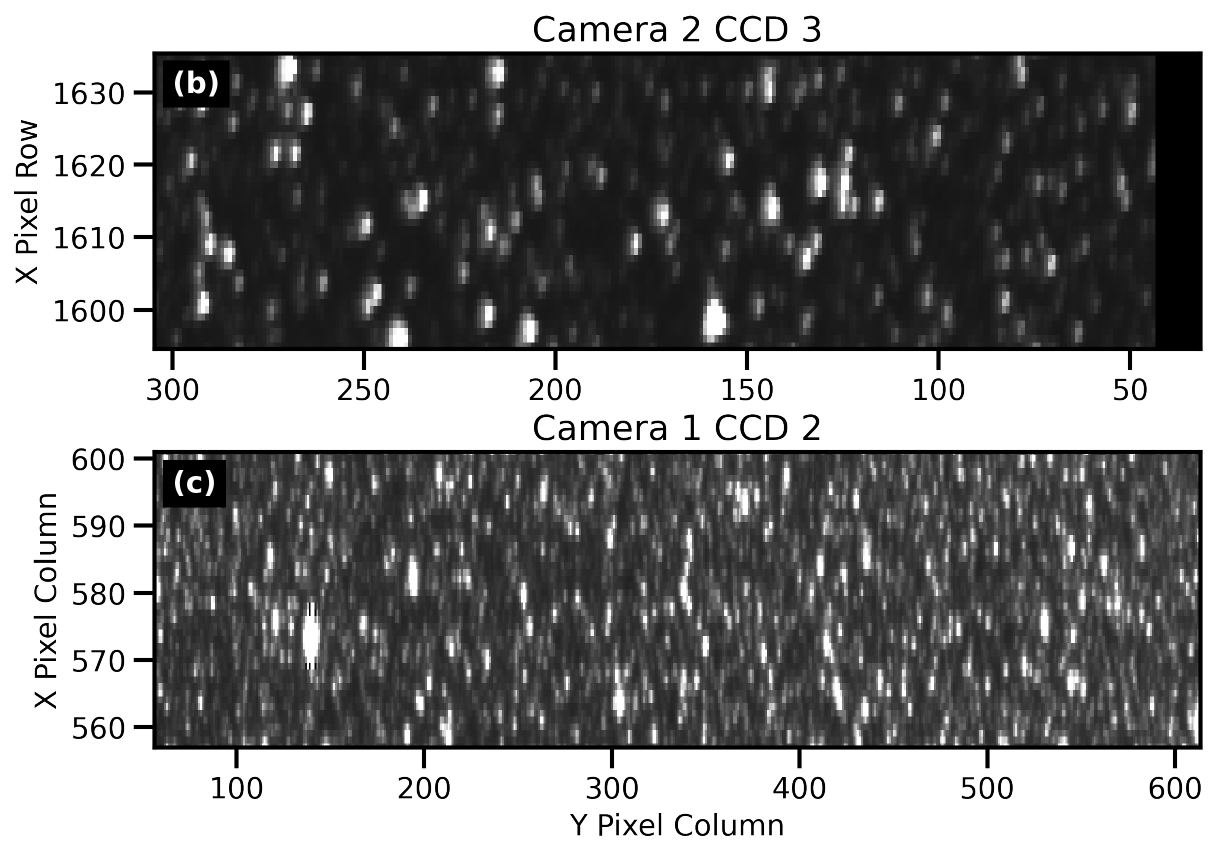}
\caption{The calibrated \textit{TESS} FFIs. (a) An example exposure of a full FFI for Camera 2 CCD 3, left, and Camera 1 CCD 2, right. We highlight the region where 3I/ATLAS was observed on both detectors by the orange box. (b) A zoom-in of the field in Camera 2 CCD 3. We note that columns $0-44$ are serial register columns and are not used in our analysis. (c) A zoom-in of the field in Camera 1 CCD 2. 3I/ATLAS enters a much more crowded field, closer to the ecliptic plane, throughout the sector. All images share the same color scaling. \href{https://github.com/afeinstein20/atlas-tess/blob/main/scripts/figure1.py}{\githubicon}
\label{fig:ffi}}
\end{figure}

Each of \textit{TESS}'s four cameras are divided into four charged-coupled devices (CCDs). Due to its high velocity, 3I/ATLAS was observed on two Camera-CCD pairings throughout the sector: Camera 2 CCD 3 and Camera 1 CCD 2. These are located next to each other in the orientation presented in Figure~\ref{fig:ffi}. Each \textit{TESS} FFI has an exposure time of $200$~seconds. The data is downloaded every $\sim 7$~days from the spacecraft due to the increased data volume of the higher cadence FFIs. These downlink times occur near apogee and perigee of the orbit of the spacecraft and thereby result in short gaps in the data. The total duration of science data collected during Sector 92 is 26.05~days. 

\subsection{FFI Selection and Background Subtraction}

We downloaded all calibrated FFIs for Camera 2 CCD 3 and Camera 1 CCD 2 using the bulk download scripts provided by the Mikulski Archive for Space Telescopes (MAST)\footnote{
\url{https://archive.stsci.edu/missions-and-data/tess}}. 3I/ATLAS was observed on pixels $x = [1592, 1635]; y=[0, 305]$ on Camera 2 CCD 3 and pixels $x = [556, 601]; y=[39,614]$ on Camera 1 CCD 2. We highlight example images of these two regions in Figure~\ref{fig:ffi}. It is important to note that columns $0-44$ are serial register columns, columns $2092-2136$ are virtual columns, and rows $2048-2078$ are buffer, smear, and virtual rows; none of these regions can be used for science.

\begin{figure}[ht!]
\includegraphics[width=1.\linewidth]{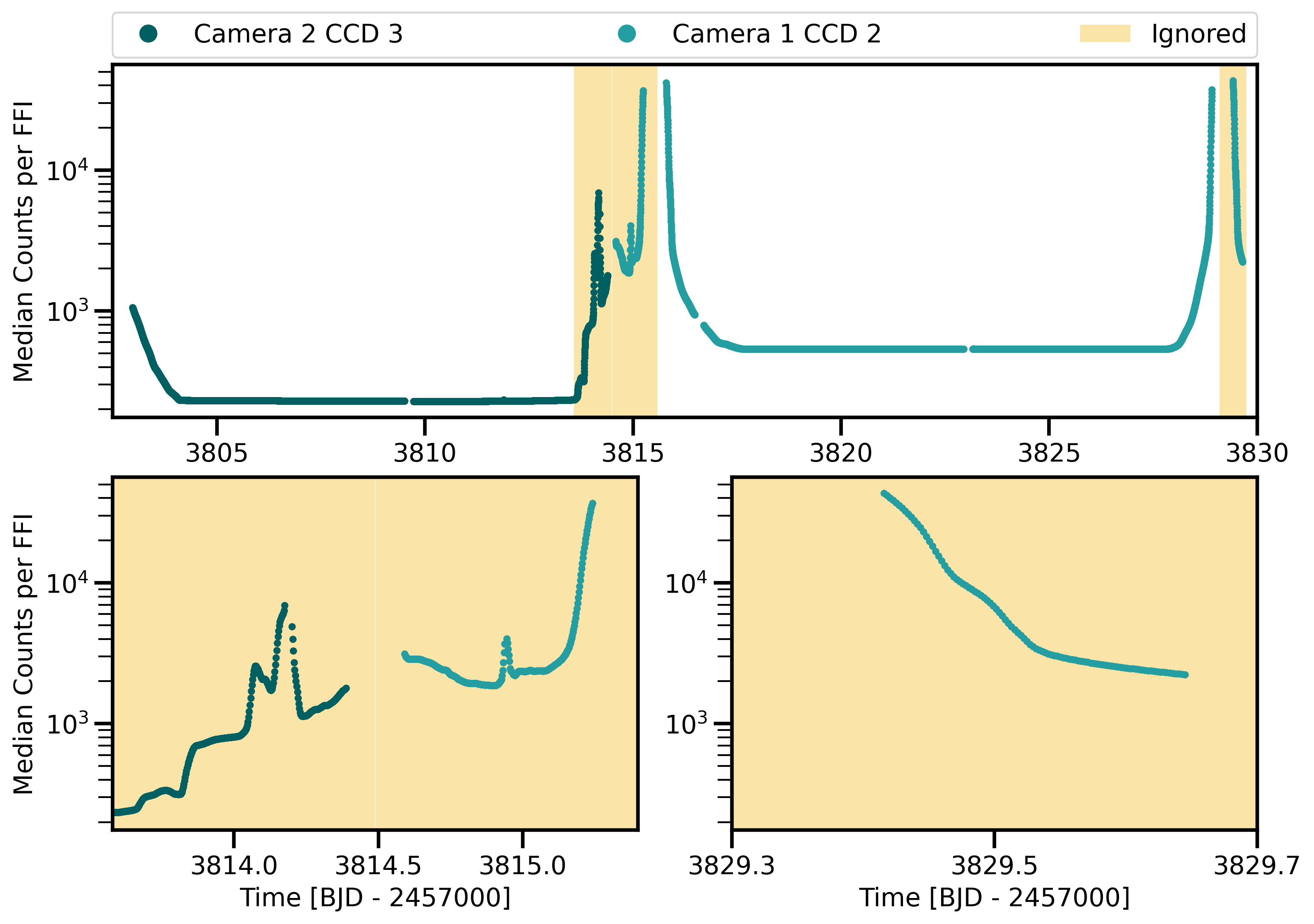}
\caption{The calibrated median counts per FFI when 3I/ATLAS is within the FOV. We highlight the strong orbital ramps, gaps, and systematics which were present in Sector 92. We choose to remove FFIs which demonstrated sharp variability and/or had unexpected gaps. These frames were visually identified and are highlighted in the bottom two subplots. FFIs not used in this analysis are highlighted in yellow. \href{https://github.com/afeinstein20/atlas-tess/blob/main/scripts/figure2.py}{\githubicon}
\label{fig:ffi_rmv}}
\end{figure}

The \textit{TESS} FFIs are contaminated by both zodiacal light and scattered earthshine contamination. In particular, the scattered earthshine manifests as strong ramps at the beginning and end of each \textit{TESS} orbit ($\sim 13$~days). While the background typically varies smoothly, there have been instances of more sporadic and variable behavior (Figure~\ref{fig:ffi_rmv}). To identify highly contaminated FFIs, we calculated the median value in each FFI . The FFIs containing highly non-smooth variability, as identified by-eye, are entirely removed from our analysis (Figure~\ref{fig:ffi_rmv}). This resulted in the removal of 9\% of the available FFIs from Camera 2 CCD and 5\% of the available FFIs from Camera 1 CCD 2. In total, after these data-quality cuts, we analyzed 9837 FFIs (353~GB). We provide more specific details pertaining to each Camera/CCD configuration in Table~\ref{tab:tess}.

We fit and subtract the background from all remaining FFIs after the heavily contaminated ones were removed. Working with the entire FFI is unnecessary. Instead, we create a ``postcard'' cutout \citep{eleanor} of the entire region where 3I/ATLAS is visible in the FFI (orange boxes in the upper two panels and full bottom two panels in Figure~\ref{fig:ffi}). Previous search and recovery analyses of known solar system bodies with \textit{TESS} have fit pixel variability using an n\textsuperscript{th}-order polynomial \citep[e.g.][]{kiss25}. We experimented with two methods to remove background contamination from these observations. First, we fit a 2\textsuperscript{nd}-order polynomial to the background, similar to \cite{kiss25}. We fit each continuous section of the light curve with these models, i.e. we chunk the observations based on gaps in the \textit{TESS} data. This leaves a total of five different segments that are fit (two segments for Camera 2 CCD 3; three segments for Camera 1 CCD 2). This experiment demonstrated that a 2\textsuperscript{nd}-order polynomial fits the baseline well, but fails to fit the orbital ramps (Figure~\ref{fig:polynomial}).

Second, we fit the background with a data-driven Savitsky-Golay filter with a pre-defined window length and a 2\textsuperscript{nd}-order polynomial. We repeat the same process of chunking the light curve based on gaps in the observations. We test several window lengths ranging from 31 - 901~pixels. The residuals of several example model fits are presented in Figure~\ref{fig:polynomial}. This experiment revealed that smaller window sizes removed astrophysical signals, while larger windows did not capture the variability of the orbital ramps. Therefore, we settled on using an average window length of 307 pixels.

\begin{table}
\centering
\caption{The locations of 3I/ATLAS and corresponding \textit{TESS} Camera/CCD times and pixel locations throughout Sector 92. We denote the number of analyzed FFIs, $N_\textrm{FFIs}$, per Camera/CCD. We note that the \textit{TESS} BJD [TBJD] time is given in units of BJD - 2457000. We denote the heliocentric and geocentric distances as $d_\odot$ and $d_\textrm{TESS}$, respectively. We denote our derived \textit{TESS} magnitude, $T_\textrm{mag}$, visual magnitude, $V$, and absolute visual magnitude, $H_V$.}
\begin{tabular}{p{1.9cm} p{2.8cm} p{2.8cm}} 
\hline
\hline
Camera/ CCD & 2/3 & 1/2 \\
\hline
Start [TBJD]& 3802.975926 & 3815.799580 \\
End [TBJD]  & 3813.474497 & 3828.911416 \\
Start [UTC] & 2025-05-07 11:24:43 & 2025-05-20 07:10:47\\
End [UTC]   & 2025-05-17 23:22:39 & 2025-06-02 09:51:49\\
x pixel range & [1592, 1635] & [556, 601]\\
y pixel range & [0, 305] & [39, 614]\\
$d_\odot$ [au] & 6.35 - 5.99 & 5.92 - 5.47 \\
$d_\textrm{TESS}$ [au] & 6.36 - 6.02 & 5.92 - 5.48 \\
$N_\textrm{FFI}$ & 4399 & 5438 \\
\hline 
$T_\textrm{mag}$ & $20.83 \pm 0.05$ & $19.28 \pm 0.05$\\
$V$ & $21.63 \pm 0.35$ & $20.08 \pm 0.35$ \\
$H_V$ & $13.72 \pm 0.35$ & $12.52 \pm 0.35$ \\
\hline
\hline
\end{tabular}
\label{tab:tess}
\end{table}

\subsection{Removing Crowded Frames}\label{sec:crowded}

\begin{figure}[ht!]
\includegraphics[width=1.0\linewidth, trim={0.1cm 0 0 0}, clip]{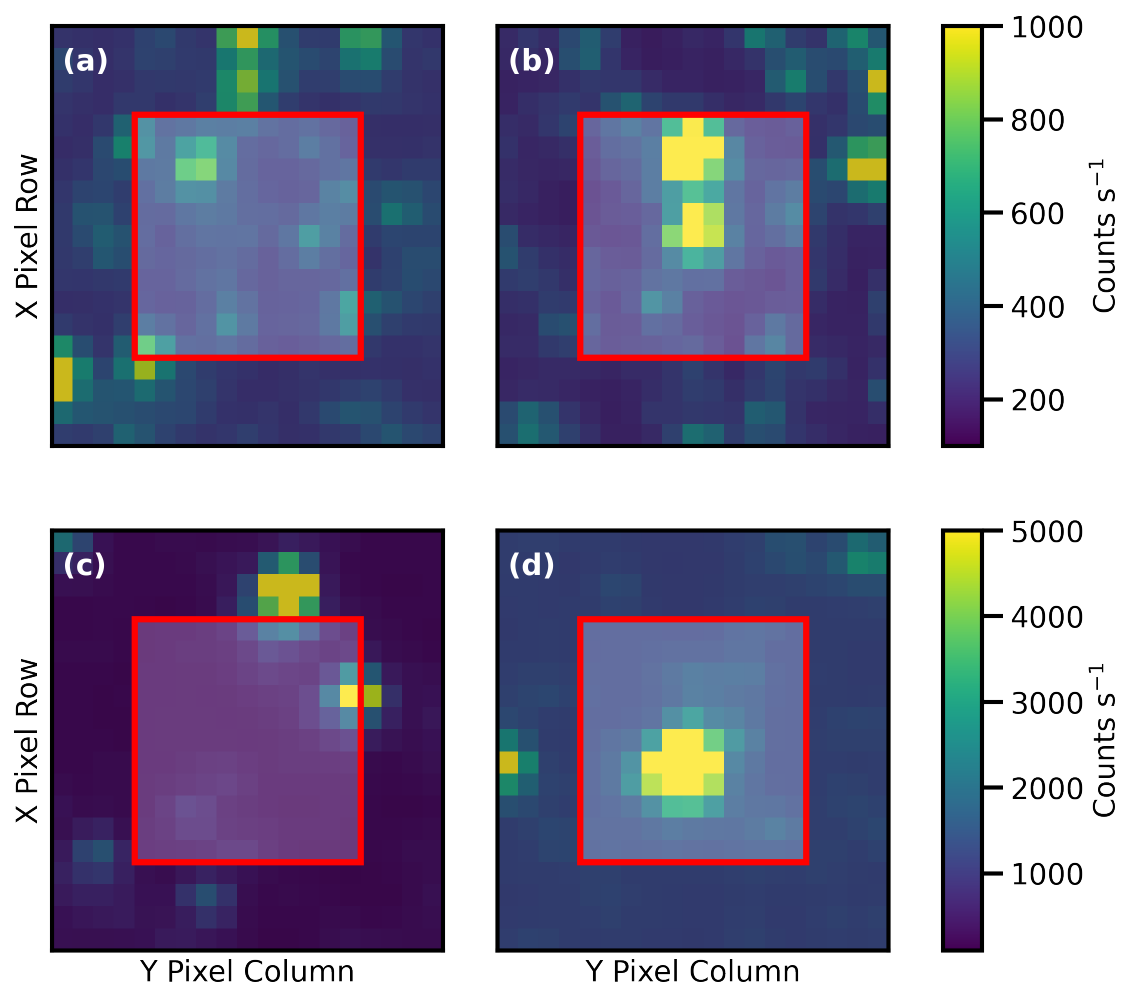}
\caption{Examples of calibrated FFI cutouts that are included (a/c) and excluded (b/d) from our deepstack image. The first and second row contains images from Camera 2 CCD 3 and Camera 1 CCD 2 respectively. The red box highlights the region we search to remove crowded images. We take this interior cutout and run it through \texttt{astropy.stats.sigma\_clip} using $\sigma = 2$. We mark images as bad when over half of the pixels within this box are $>2\sigma$ above the average value. This sufficiently removes images where the path of 3I/ATLAS crosses background sources.
\href{https://github.com/afeinstein20/atlas-tess/blob/main/scripts/figure3.py}{\githubicon}
} 
\label{fig:removed}
\end{figure}

Due to the crowdedness in some of the images, we create a filter to remove images which may have significant stellar contamination. The final deepstack images in this work  only include frames that are not significantly contaminated.  The contamination criteria  is defined as the following: Because we are only interested in crowding near the location of 3I/ATLAS, we only consider sources within an $11 \times 11$~window around the central location. We use \texttt{astropy.stats.sigma\_clip} to determine how many pixels are $> 2\sigma$ outliers compared to the mean value. We assume the default settings for \texttt{astropy.stats.sigma\_clip}, other than defining the $\sigma$ value. We mask any frame for which more than half of the pixels within this region are flagged as $>2 \sigma$ outliers.   Examples of non-crowded versus crowded frames  are  shown in Figure~\ref{fig:removed}.

\subsection{Deep Image Creation}

After the background models are removed, we employ a shift-stack algorithm to create a \textit{TESS} deepstack image. We choose not to combine images across detectors, as they may have different systematics. Our algorithm works as follows: (I) Query the location of 3I/ATLAS from the JPL Horizons Small Bodies Database for each FFI; (II) Use the world coordinate system information to convert the queried (RA, Dec) to the (x,y) pixel location of 3I/ATLAS. We convert the (x, y) location from float values to integers using \texttt{np.round}, in order to accurately determine the nearest pixel; (III) Create a $17 \times 17$~pixel cutout centered on the integer (x, y) coordinates where 3I/ATLAS is expected to be; (IV) Repeat this procedure for all integrations; (V) Sum all of the images. Additionally, due to the apparent location with respect to the galactic plane, we create a data quality flag to track particularly crowded fields. The shift-stack technique implemented in this paper works well for objects with known orbital parameters, but may be more computationally inefficient when completing a blind-search for new objects. However, it has been employed with some success for example by \citet{Fraser2024} --- albeit using  more sensitive data. A coarse summation presented here is sufficient for \textit{TESS} due to its large pixel scale.

\begin{figure}[ht!]
\includegraphics[width=1.0\linewidth]{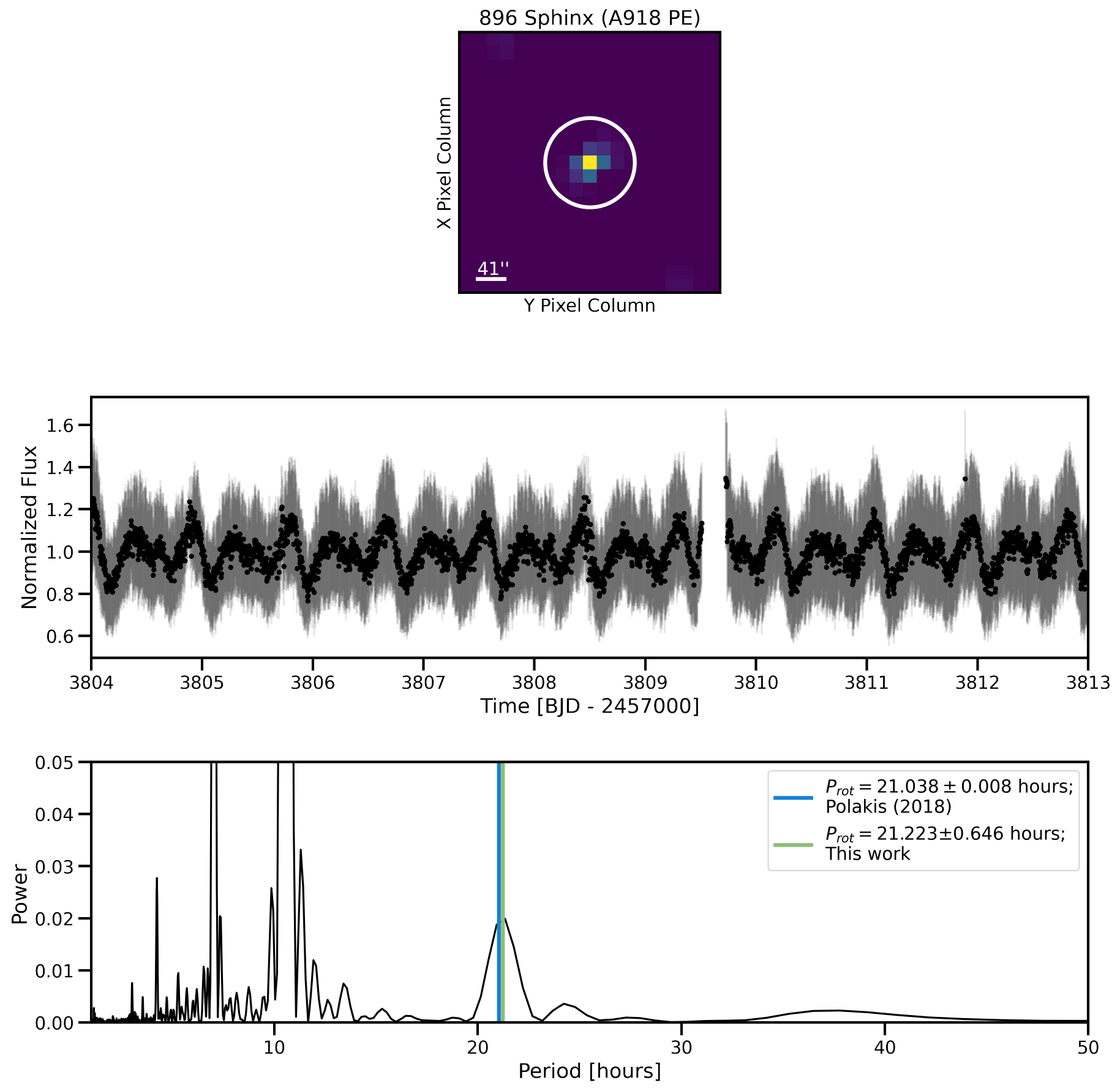}
\caption{Recovery of 896 Sphinx (A918 PE). We use this object  ($H_V = 15$) as a demonstration of our background subtraction and shift-stack technique. (a) The median background-subtracted deepstack image of the main belt minor planet 896 Sphinx. 896 Sphinx is highlights in the image by a white circle. We provide the \textit{TESS} pixel scale in the bottom left. (b) The extracted light curve of 896 Sphinx, which shows clear variability on an hours-long timescale. (c) The associated periodogram. 896 Sphinx has a measured rotation period of $21.038 \pm 0.008$~hours \citep{polakis18}. We find a best-fit period of this object from \textit{TESS} is $21.858 \pm 0.577$~hours, which is with $1.5\sigma$ agreement with \citep{polakis18}. The discrepancy between periods is likely due to the lack of \textit{TESS} sensitivity of faint objects. These images were stacked from Camera 2 CCD 3.  \href{https://github.com/afeinstein20/atlas-tess/blob/main/scripts/figure4.py}{\githubicon}}
\label{fig:asteroid}
\end{figure}

\subsection{Recovery of Known Minor Planet 896 Sphinx}

We validated our algorithm by recovering another small body which was observed in the same field as 3I/ATLAS. We find that the known main-belt minor planet 896 Sphinx (A918 PE) (H$_V\simeq15$) was observed on Camera 2 CCD 3. 896 Sphinx was initially discovered in 1918 by M. Wolf. It has a known diameter of 11.9~km \citep{masiero14} and rotation period of $21.038 \pm 0.008$~hours \citep{polakis18}. We apply our background-correction and shift-stack method outlined above to the region where 896 Sphinx is within the FFIs.

We present the deepstack image of 896 Sphinx in Figure~\ref{fig:asteroid}. We find that we are able to recover 896 Sphinx at $6.6 \sigma$. Additionally, we extract a light curve using a $3 \times 3$-pixel mask centered on 896 Sphinx. We present the extracted normalized light curve and associated errors in Figure~\ref{fig:asteroid}. We run this light curve through a Lomb-Scargle periodogram \citep{lomb, scargle} to recover the known rotation period of the object. We find that periods $P < 10$~hours are dominated by instrumental noise. We ignore these periods. We find the peak of max power for $P > 10$~hours. We fit this peak with a Gaussian function and optimize our fit using a non-linear least squares approach. We find that the peak occurs at $21.858 \pm 0.577$~hours. Our error is adopted from the width of the best-fit Gaussian. Our recovered period is within $1.5\sigma$ of the period presented in  \cite{polakis18}. The discrepancy between our rotation period and the archival rotation period may be due to the lack of sensitivity of \textit{TESS} for faint objects.

\begin{figure}[ht!]
\includegraphics[width=1.0\linewidth]{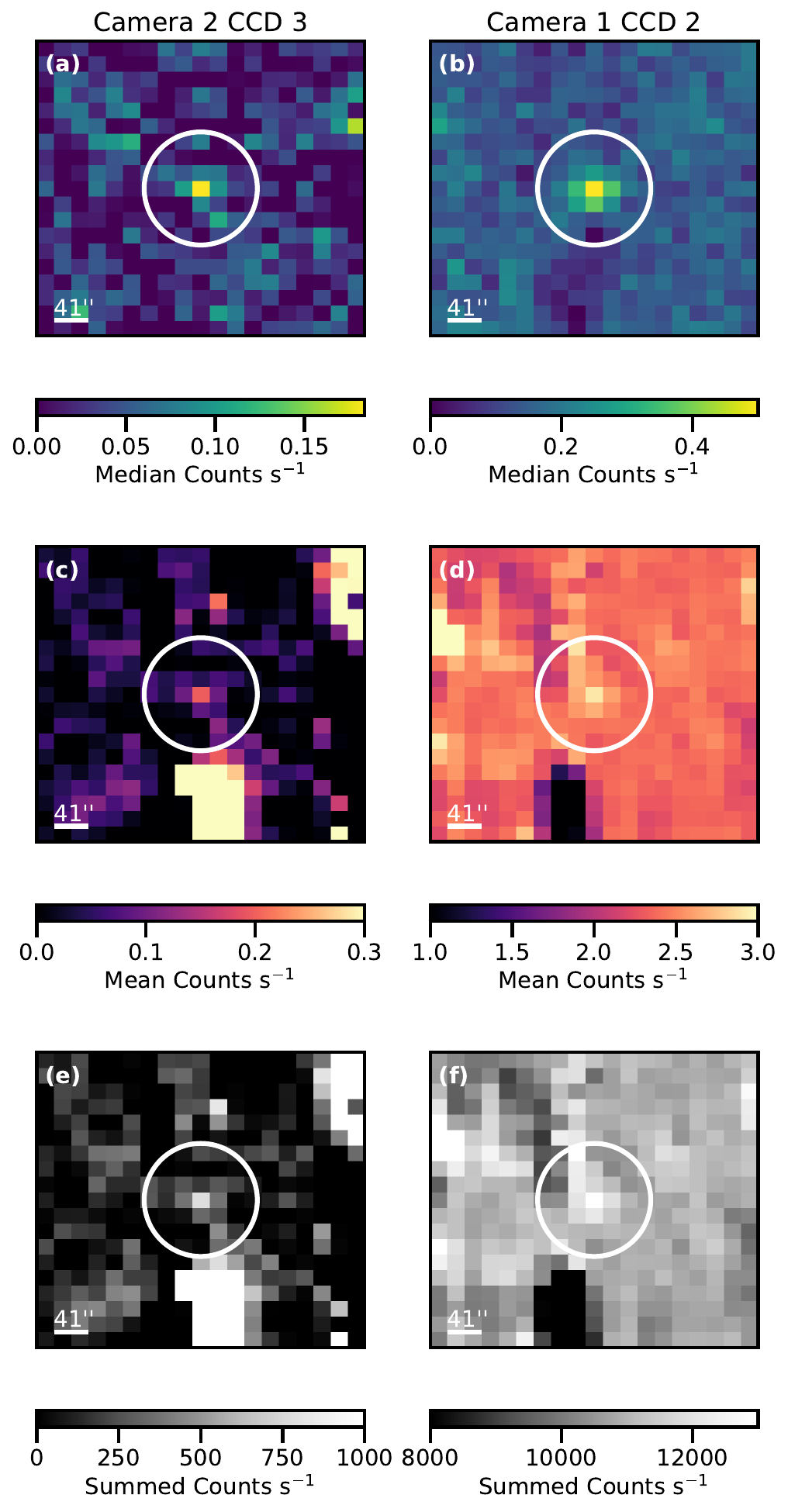}
\caption{Deepstacked background-subtracted images of 3I/ATLAS in the \textit{TESS} FFIs. We present two images per Camera/CCD pairing, labeled at the top of the columns. Panels a/b, c/d, and e/f are the median, mean, and summed stacked images, respectively. 3I/ATLAS is centered in each image and is highlighted with a white circle. We limited the stacked FFIs to cutouts which are not overcrowded with bright sources, as defined in Section~\ref{sec:crowded}. Due to the large pixel scale of \textit{TESS} (bottom left), we are unable to resolve any astrometric information from the images. \href{https://github.com/afeinstein20/atlas-tess/blob/main/scripts/figure5.py}{\githubicon}
\label{fig:stacked}}
\end{figure}

\subsection{Recovery of 3I/ATLAS}

Figure~\ref{fig:stacked} shows the background-removed deepstack images and associated errors of 3I/ATLAS as observed by \textit{TESS}. These deepstacks are limited to cutouts which are not overcrowded with bright sources, as defined in Section~\ref{sec:crowded}. The median, mean, and summed images are shown for both detectors. The detection threshold for 3I/ATLAS is calculated by propagating errors from the calibrated FFIs. Although the median deepstack images are implemented for the remainder of this paper, it should be noted that 3I/ATLAS is visible by-eye in all three types of images in Figure~\ref{fig:stacked}. The central pixel in  Camera 2 CCD 3 has a median counts value of $0.173 \pm 0.004$ e\textsuperscript{-} s$^{-1}$. The background scatter is calculated as the median and standard deviation of the singular pixels surrounding the central pixel exercise. The background has an averaged count value of $0.022 \pm 0.008$ e\textsuperscript{-} s$^{-1}$. Therefore, we are able to recover 3I/ATLAS at $19\sigma$ in this summed deepstack image. The same exercise is repeated for Camera 1 CCD 2. The central pixel in this image has a summed counts value of $0.589 \pm 0.016$ e\textsuperscript{-} s$^{-1}$. The background has an averaged count value of $0.141 \pm 0.039 $ e\textsuperscript{-} s$^{-1}$. Therefore, we are able to recover 3I/ATLAS at $11\sigma$ in this summed deepstack image. It is not surprising that we are unable to recover 3I/ATLAS at the same high fidelity as the previous image. This is because the field becomes increasingly more crowded as 3I/ATLAS' apparent motion places it closer to the galactic plane.

\section{Results}\label{sec:discuss}

3I/ATLAS has displayed visible cometary activity months prior to perihelion.  It is feasible  that the level of activity has changed  and will continue to change throughout its orbit.  Light curves from images when 3I/ATLAS was at further heliocentric distances could provide insights into variations in the activity and nucleus properties such as the rotation period.   In this section, we therefore extract the 20-day \textit{TESS} light curve and calculate the resulting secular light curve of 3I/ATLAS.

\subsection{Magnitude Calculation}

We use a $3 \times 3$ pixel aperture on the median deepstacked image to calculate the magnitude of 3I/ATLAS. We convert the calibrated flux in e$^-$ s$^{-1}$ to \textit{TESS} magnitude using the following relationship:

\begin{equation}\label{eq:tmag}
    T_\textrm{mag} = -2.5\textrm{log}_{10}(c) + 20.44\,.
\end{equation}

In Equation \ref{eq:tmag}, $c$ is the number of counts per second. This conversion is provided in the \textit{TESS} Instrumental Handbook \citep{vanderspek18}. We adopt the error on the zeropoint of $0.05$~mag. We note that while the zeropoint may change slightly between sectors, cameras, and/or CCDs, it is standard practice to not recalculate the value for each observation combination \citep[e.g.][]{farnham2019_46ptess, fausnaugh23, andrews25}. We calculate $T_\textrm{mag}$ using a $3\times3$ aperture on the median deepstack images presented in Figure~\ref{fig:stacked}. For Camera 2 CCD 3, we derive $c_{23} = 0.699 \pm 0.004$~e\textsuperscript{-} s$^{-1}$. For Camera 1 CCD 2, we derive $c_{12} = 2.904 \pm 0.016$~e\textsuperscript{-} s$^{-1}$.

We find that 3I/ATLAS has a $T_\textrm{mag} = 20.83 \pm 0.05$ in the Camera 2 CCD 3 image and a $T_\textrm{mag} = 19.28 \pm 0.05$ in the Camera 1 CCD 2 image. Our calculated $T_\textrm{mag}$ is in agreement with \cite{martinez25}. Examining the origins of slight offsets between our extracted values is beyond the scope of this work. These two magnitudes are averaged over the two observing windows in Table \ref{tab:tess}, which span a range in heliocentric distance of 6.4 to 5.4 au, respectively. In that same time period the distance between 3I/ATLAS and the TESS spacecraft decreased by 0.9 au. The rapid change in both these properties can be used to constrain any changes to the overall activity of 3I/ATLAS between both times.

From geometric effects alone, and ignoring the small ($\sim2.5^{\circ}$) phase angle change, we would expect the flux from 3I to increase by a factor of 1.5. Converting this back into magnitudes we expected 3I/ATLAS to have a measured magnitude of 20.5 in the Camera 1 CCD 2 images, rather than the observed $T_\textrm{mag}$ of 19.28, indicating a factor of 5 increase in the flux, rather than 1.5. Within our uncertainty on the \textit{TESS} magnitude, our observations are statistically inconsistent with an asteroid-like reflectance model that ignores dust production.

\subsection{Photometric Light Curve}

\begin{figure}[ht!]
\includegraphics[width=1.0\linewidth, trim={0.1cm 0 0 0}, clip]{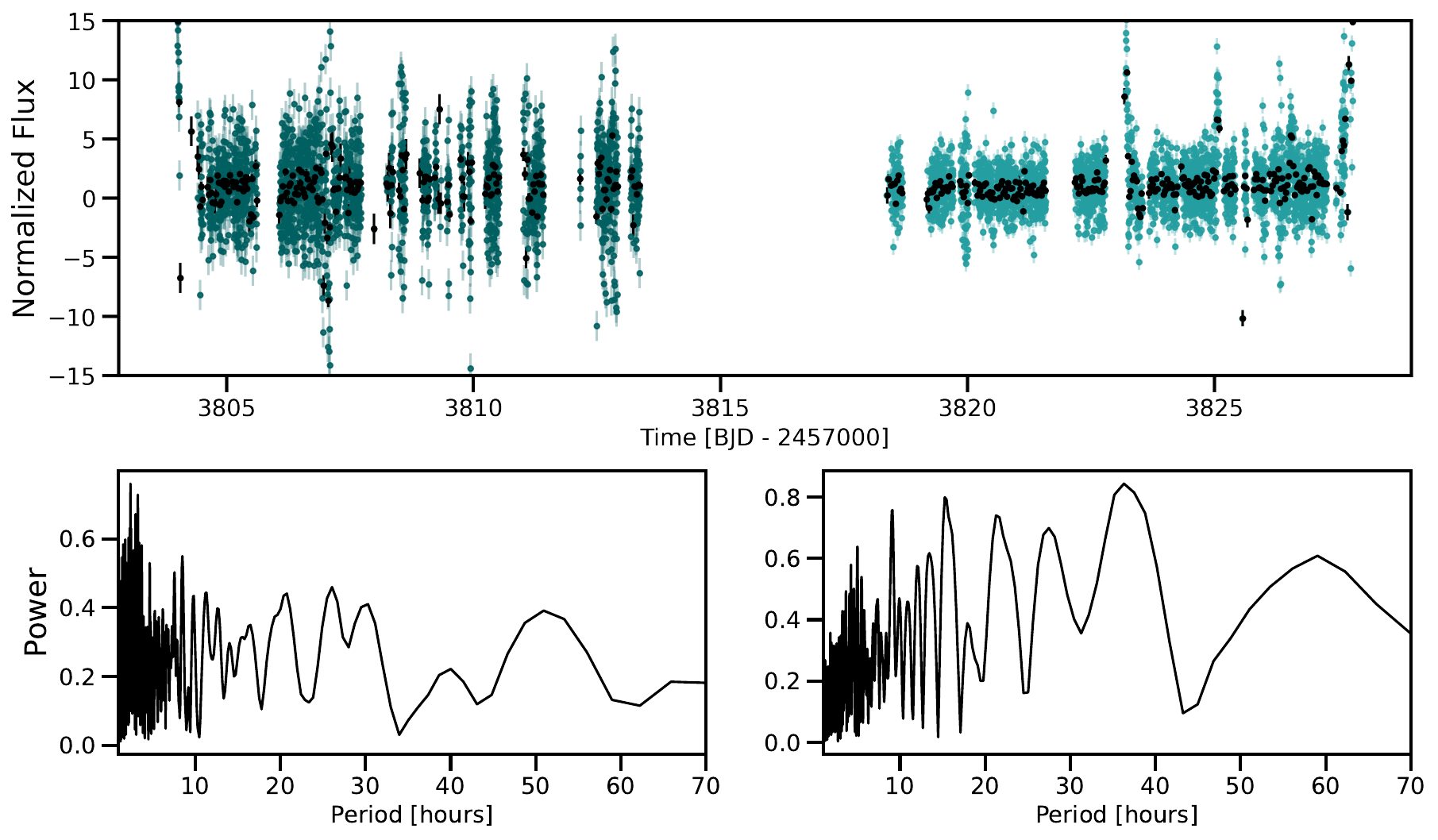}
\caption{Compiled normalized \textit{TESS} light curve of 3I/ATLAS and associated error bars. We bin the light curve to $\sim 36$~minutes in black. These \textit{TESS} data offer an advantage when trying to measure the nucleus' rotation period, as 3I/ATLAS was farther from perihelion. We run the binned light curve through a Lomb-Scargle periodogram (bottom panels). We do not find similarities in periods between both light curves. \href{https://github.com/afeinstein20/atlas-tess/blob/main/scripts/figure6.py}{\githubicon}
\label{fig:lc}}
\end{figure}

We extract a light curve using the same $3\times3$-pixel aperture and propagate the errors from the calibrated FFIs. We normalize the light curve. Additionally, we bin our light curve to 36-minute bins. We analyze the binned light curve with a Lomb-Scargle periodogram to search for evidence of periodicity. We do this for each light curve independently. We search for a rotation period between $P_\textrm{rot} = 1-70$~hours, which is mostly consistent with $P_\textrm{rot}$ of other kilometer scale small bodies in the solar system \citep{Warner2009}. Both of the light curves and affiliated periodograms are presented in Figure~\ref{fig:lc}.

\begin{figure}[ht!]
\includegraphics[width=1.0\linewidth, trim={0.1cm 0 0 0}, clip]{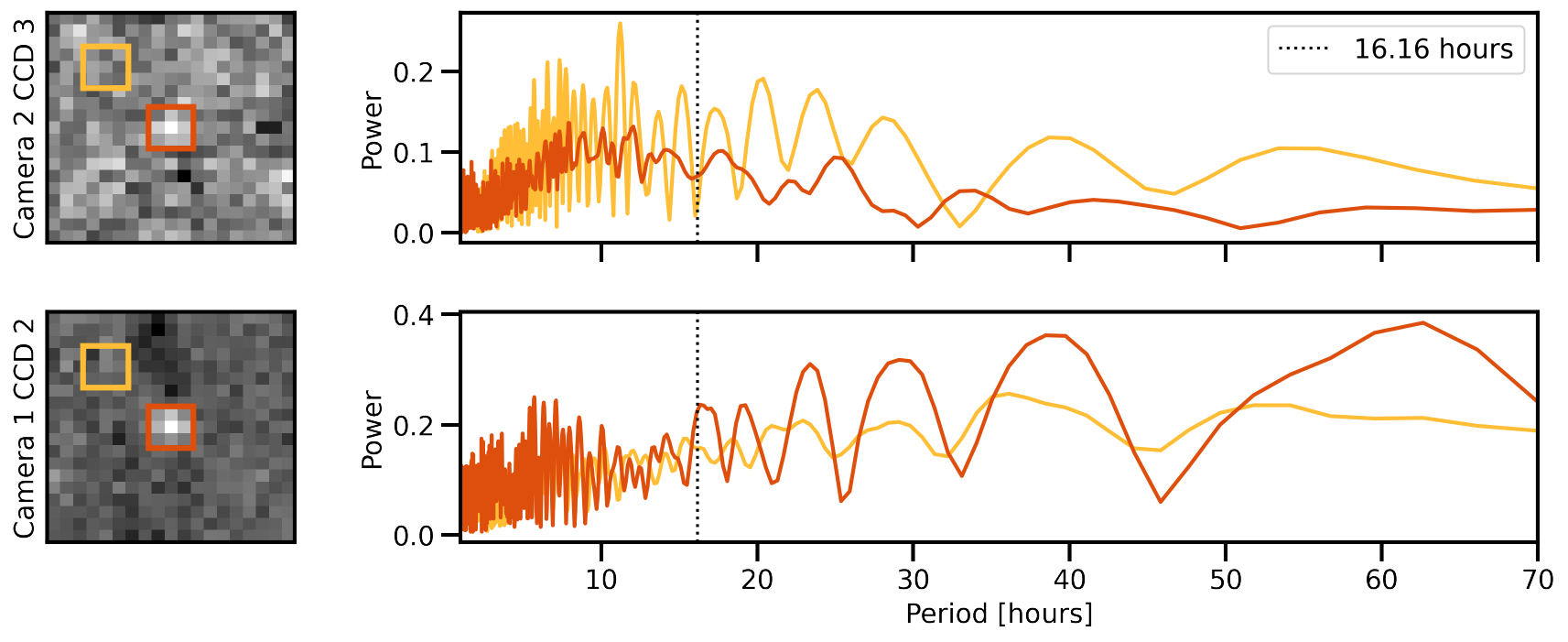}
\caption{Comparison of the Lomb-Scargle periodogram for 3I/ATLAS (orange) and a background pixels (yellow). We select a pixel in the same location across both images. We mark the recently hypothesized rotation period of 3I/ATLAS as a horizontal dotted line \citep{Marcos2025, santanaros25}. The background pixels here are representative of systematics seen across all pixels within a $5\times5$ region around the central pixel. We find that the light curve and resulting periodogram are dominated by noise and therefore we cannot constrain the rotation period of the nucleus. \href{https://github.com/afeinstein20/atlas-tess/blob/main/scripts/figure7.py}{\githubicon}}
\label{fig:comp_ls}
\end{figure}

There are several similarities between the Lomb-Scargle periodograms. First, we find that  $P_\textrm{rot} < 10$~hours are dominated by noise and/or unremoved systematics. Second, we see that there is no agreement between periodograms at $P > 40$~hours. Third, we see a some strong periodicity between 10-30~hours. This is more apparent in the light curve from Camera 1 CCD 2, when 3I/ATLAS 0.4~au closer. In particular, we find a peak in the periodogram close to the recently recovered period of 16.16-16.79~hours presented in \cite{Marcos2025, santanaros25}.

To further assess this potential rotational signature, we run the same Lomb-Scargle test for a nearby background pixels in the FFI. Because the background can rapidly vary across the \textit{TESS} detector, we select an example background pixel within $5$-pixels of 3I/ATLAS. The results of this test are presented in Figure~\ref{fig:comp_ls}. We find that the background pixel periodogram for Camera 2 CCD 3 shows a similar peak at $\sim 16$~hours, but we do not see a similarly strong peak around this time for Camera 1 CCD 2. However, there are stronger periodic peaks in this periodogram, so it is challenging to assess whether or not this 16~hour period is physically motivated by the \textit{TESS} observations alone. 

\begin{figure}[ht!]
\includegraphics[width=1.0\linewidth, trim={0.1cm 0 0 0}, clip]{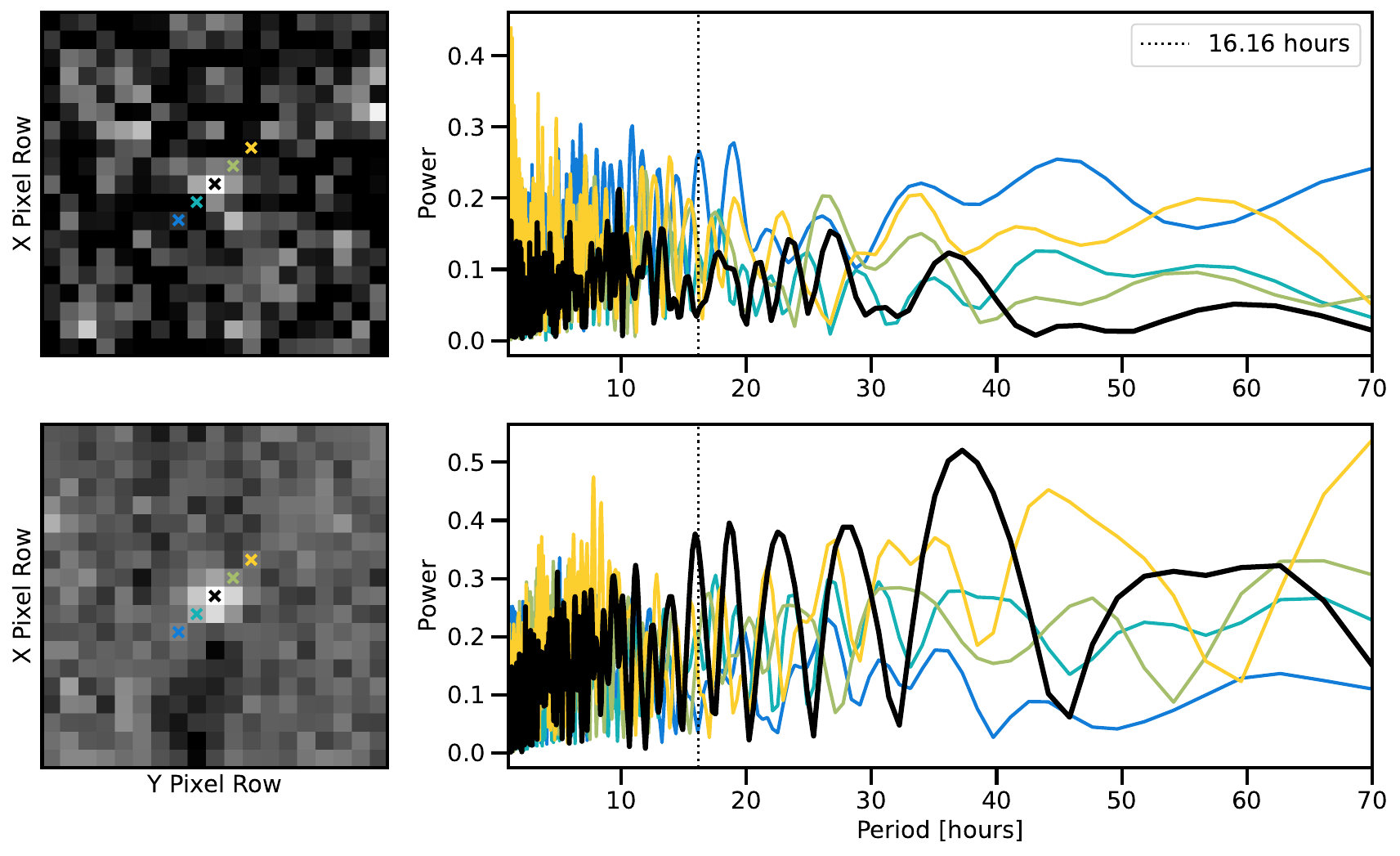}
\caption{Comparison of the Lomb-Scargle periodogram for 3I/ATLAS (black) and trailing/leading pixels (colored). The pixels and corresponding periodograms are marked in the same color. The top/bottom row is for the observations from Camera 2 CCD 3/Camera 1 CCD 2. There is a strong periodic signal at 33~hours in the trailing and leading pixels,  offset from the peak from 3I/ATLAS by $\sim 4$~hours. Moreover, these pixels all display some periodicity at $P = 20-30$~hours, comparable to that seen in 3I/ATLAS. There are similarities in the periodograms between 3I/ATLAS and the trailing/leading pixels such as the signals at $24-27$~hours and at $35-41$~hours. These similarities indicate that  significant rotational information regarding 3I/ATLAS cannot be extracted from the \textit{TESS} light curves. \href{https://github.com/afeinstein20/atlas-tess/blob/main/scripts/figure8.py}{\githubicon}
\label{fig:trail_lead}}
\end{figure}

We complete this exercise for all of the pixels within a $5\times5$ region around 3I/ATLAS. We pay particular attention to pixels that lead and trail the source on sky, as they should reveal systematics produced by nearby stars. The leading and trailing pixels are defined as those  diagonal to the center from the bottom left to the upper right (the direction of motion of 3I/ATLAS). The   light curve extracted from each pixel is binned at a 36~minute resolution. A Lomb-Scargle periodogram is performed on each resulting light curve across the same frequency range (Figure~\ref{fig:trail_lead}). There are some similarities between the leading/trailing pixels that do \textit{not} align with the periodic signals observed from 3I/ATLAS. In particular,  the trailing/leading pixels display a similar periodic signal at 29~hours in the data from Camera 2 CCD 3. This signal  is offset from a peak in the 3I/ATLAS observations by $\sim 4$~hours. There are similar offsets between the trailing/leading and 3I/ATLAS pixels for Camera 1 CCD 2.

Given the similarities in noise properties of the periodograms, it is challenging to draw any confident conclusions about 3I/ATLAS. Therefore, we conclude that these \textit{TESS} light curves do no definitively capture a $P_\textrm{rot}$ from the nucleus. This may be due to ongoing outgassing even at large heliocentric distances or due to the sensitivity of \textit{TESS}, making any possible magnitude changes from the baseline difficult to definitively measure.

\subsection{Secular Light Curve}

We calculated the inferred absolute V magnitude of 3I/ATLAS from these observations to add to the light curve presented in \cite{seligman25}. The secular light curve of a comet provides insight into the type of volatile driving activity --- or at least the volatility of the species driving activity. For example, \citet{womack2021} presented a 4 year secular light curve of  C/1995 O1 (Hale-Bopp) to infer that CO was responsible for the activity exterior to 2.6-3.0 au, while H$_2$O was responsible for it interior to that.   

We convert the calculated \textit{TESS} magnitude to absolute V magnitude, $H_V$, using the following. \cite{farnham2021_un271tess} derived a rough relationship between the apparent visual magnitude, $V$, and \textit{TESS} magnitude of 

\begin{equation}\label{eqn:apparent}
    V = T_\textrm{mag} + 0.8\,.
\end{equation}
This relationship was derived using the Web TESS Viewing Tool assuming typical comet colors derived from the average of several Jupiter Family comets, long-period comets, and active centaurs. This relationship has an uncertainty of $\pm 0.3$~mag, which is dominated by the uncertainty in comet colors. We convert $V$ to $H_V$ using the following:

\begin{equation}\label{eqn:absolute}
    H_V = V - 2.5\,n \,\textrm{log}_{10}(d_\odot) - 5 \,\textrm{log}_{10}(d_\textrm{TESS})\,.
\end{equation}
In Equation \ref{eqn:absolute}, $d_\odot$ is the comet-Sun distance, $d_\textrm{TESS}$ is the comet-\textit{TESS} distance, and $n$ is an activity index. $n$ has a typical range from $n = 2-6$, where $n=2$ assumes the body is inactive and $n=6$ assumes the object has strong activity \citep{everhart67}. Given that 3I/ATLAS is at a distance of 6.4~au in these images, we assume $n=2$.

\begin{figure}[ht!]
\includegraphics[width=1.0\linewidth, trim={0.1cm 0 0 0}, clip]{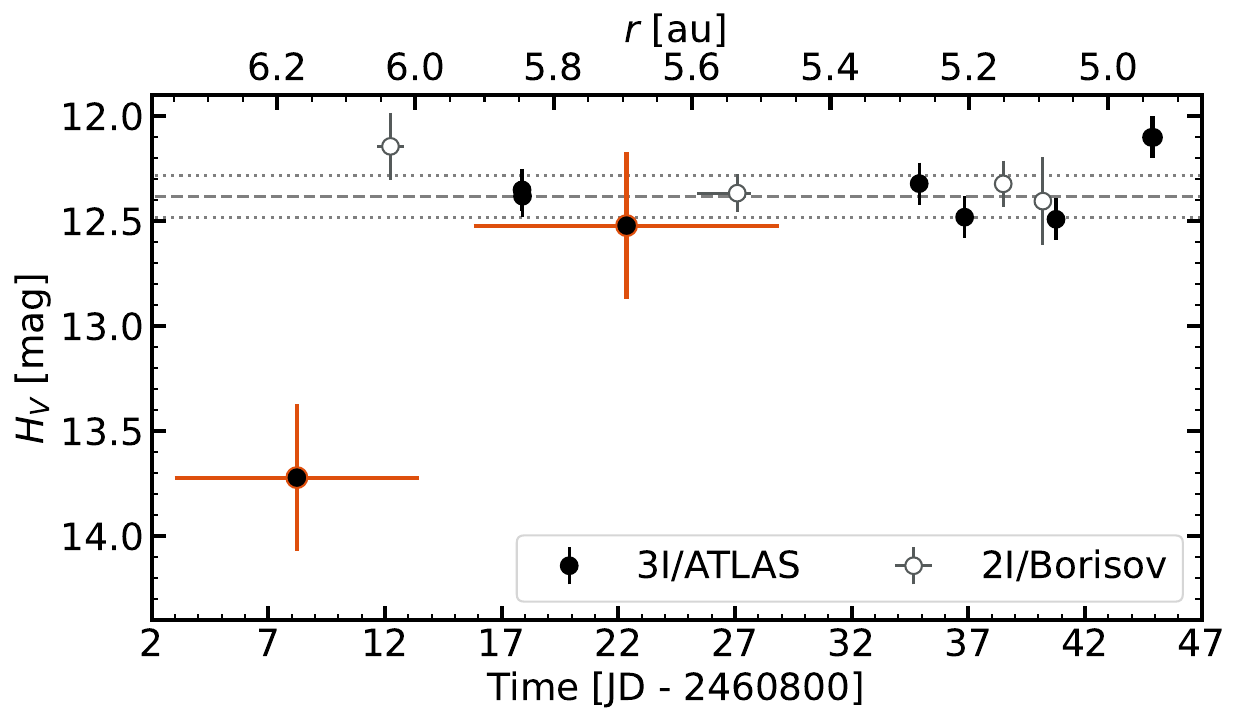}
\caption{Absolute visual magnitude, $H_V$ computed from \textit{TESS} (orange) and ZTF observations of 3I/ATLAS (black) as compared to 2I/Borisov (gray). The median and $1\sigma$ standard deviations of the ZTF observations are plotted as dashed and dotted lines, respectively. The top axes represents the heliocentric distance of 3I/ATLAS and 2I/Borisov at the time of the observation. ZTF observations of 3I/ATLAS were derived in \cite{seligman25}; pre-perihelion observations of 2I/Borisov were derived in \cite{Ye2020}. We note that the lower time x-axis is applicable to 3I/ATLAS and not to 2I/Borisov. \href{https://github.com/afeinstein20/atlas-tess/blob/main/scripts/figure9.py}{\githubicon}
\label{fig:absolute_mag}}
\end{figure}

We use the average $d_\odot$ and $d_\oplus$ for each Camera/CCD configuration presented in Table~\ref{tab:tess}.  From our calculated \textit{TESS} magnitude, following Equations~\ref{eqn:apparent} and \ref{eqn:absolute}, and adopting the uncertainty presented in \cite{farnham2021_un271tess}, we derive $H_V = 13.72 \pm 0.35$ and $12.52 \pm 0.35$. We plot discovery and precovery ZTF observations of 3I/ATLAS in Figure~\ref{fig:absolute_mag}.

\section{Discussion}\label{sec:conclusion}

We present two deepstack images and a 20-day light curve of 3I/ATLAS as observed with NASA's \textit{Transiting Exoplanet Survey Satellite}. These data were obtained from 07 May to 03 June 2025, predating the initial discovery by nearly two months \citep{Denneau2025}. From our two epochs of these data we calculate that 3I/ATLAS had a $T_\textrm{mag} = 20.83 \pm 0.05$ and $T_\textrm{mag} = 19.28 \pm 0.05$, the latter of which is consistent with more recently obtained $R_C$ observations from the TRAPPIST-North and TRAPPIST-South telescopes \citep{seligman25}. Moreover, we identify a periodic signal at $\sim 16$~hours in the light curve from Camera 1 CCD 2. However,  we find no statistically significant evidence of a rotation period from the nucleus when comparing the light curve of 3I/ATLAS to a neighboring background pixel. The lack of observed rotation may be due to the presence of a coma, uncertainties in the size of the nucleus if there is no coma, or decreased sensitivity of \textit{TESS} to faint objects. 

The derived $H_\textrm{V}$ from these data indicate that 3I/ATLAS brightened more than can be expected from the decrease in helio- and TESS-centric distances between the two periods. It is worth stating that the object traversed nearly 1 au during the course of these observations. We note that the absolute magnitude of 3I/ATLAS in our first epoch,$H_\textrm{V}$=13.9, is consistent with activity from a small (R$<$2.8 km, $H_\textrm{V}>
$15.4) nucleus derived from HST WFC3 imaging \citep{jewitt2025hubblespacetelescopeobservations}. This provides tentative evidence that 3I/ATLAS was active during the fortuitous TESS observations \textit{prior} to its discovery on UT 01 July 2025. However, the 3$\sigma$ error  technically permits the possibility that 3I/ATLAS was fainter by $\sim$1 magnitude  during these observations. This is due to the uncertainties in the conversion from $T_\textrm{mag}$ magnitudes to $V$ of 0.3 \citep{vanderspek18,farnham2021_un271tess}. Therefore, we conclude that these observations are consistent with the hypothesis that 3I/ATLAS was weakly active prior to its discovery. Future work could include deriving a new relationship between $T_\textrm{mag}$ and other bandpasses for ISOs.

This distant activity of 3I/ATLAS --- if confirmed --- could be indicative of mass loss driven by a mechanism other than the sublimation of H$_2$O ice. One possibility is that 3I/ATLAS is enriched in hypervolatiles such as CO or CO$_2$. 2I/Borisov displayed a higher production of CO than H$_2$O when these two quantities were measured contemporaneously \citep{Bodewits2020,Cordiner2020,Xing2020}, although no direct measurement of CO$_2$ has been successfully obtained in an interstellar object to date.\footnote{Production rates of CO$_2$ have been inferred from the ration of O I emission features in the visible, \citealt{mckay2024borisov_co2}}  It is also possible that this activity is driven by a more exotic hypervolatile such as H$_2$ or N$_2$. The nongravitational acceleration of 1I/`Oumuamua energetically would have required the sublimation of an ice more volatile than H$_2$O \citep{Sekanina2019,Seligman2020}. It has therefore been proposed that this acceleration was driven by  outgassing of N$_2$ \citep{Jackson2021}, CO \citep{Seligman2021}, or radiolytically produced  H$_2$ \citep{bergner2023h2}. It is possible that distant activity of 3I/ATLAS is driven by one of these, but determining which neutral is responsible requires spectroscopic confirmation or long-term lightcurve modeling \citep{bufanda2023aas}. 

The change in absolute magnitude between our two deep stacks indicates a change in activity, likely the result of a rapidly warming thermal environment. Continued searches for precovery images and extensive monitoring of 3I/ATLAS prior to perihelion will be essential for characterizing this trend. In particular, other precovery images which may include stellar occultations of appulses \citep{ortiz2023chiron,pereira2025occultation}, especially from multi-filter observations, could constrain the coma density and composition. Moreover, spectroscopic follow-up observations could also confirm the existence of hypervolatiles in 3I/ATLAS.

Observations of future ISOs could also shed light on the extent to which this population displays distant activity and is enriched with hypervolatiles. Existing all-sky surveys, including the new Rubin Observatory Legacy Survey of Space and Time (LSST) have been predicted to detect future ISOs \citep{Moro2009,Engelhardt2014,Cook2016,Seligman2018,Hoover2022,Miller2022,Landau2023,Marceta2023a,Marceta2023b,Dorsey2025}. Reorienting \textit{TESS} to observe the ecliptic for longer periods of time in future extended missions could aid in the precovery observations of newly detected ISOs.


\section{Data and Software Availability}

The calibrated \textit{TESS} FFIs are available for bulk download on the Mikulski Archive for Space Telescopes.\footnote{\url{https://archive.stsci.edu/tess/bulk_downloads/bulk_downloads_ffi-tp-lc-dv.html}} All of the analysis Python scripts, data behind the figures, and Python scripts for reproducing the figures presented here are available on GitHub.\footnote{\url{https://github.com/afeinstein20/atlas-tess}} Additional, larger, data products will be made available on Zenodo upon publication.

This work made use of the following open-source Python packages: \texttt{astropy} \citep{2013A&A...558A..33A, 2018AJ....156..123A, 2022ApJ...935..167A}, \texttt{numpy} \citep{numpy}, \texttt{scipy} \citep{2020SciPy}, \texttt{matplotlib} \citep{matplotlib}.

\section{Acknowledgments}

We thank Henry Hsieh, Davide Farnocchia, and Marco Micheli for thoughtful insights. We thank the anonymous referee for their comments which have improved the quality and clarity of this manuscript. A.D.F. acknowledges funding from NASA through the NASA Hubble Fellowship grant HST-HF2-51530.001-A awarded by STScI. D.Z.S. is supported by an NSF Astronomy and Astrophysics Postdoctoral Fellowship under award AST-2303553. This research award is partially funded by a generous gift of Charles Simonyi to the NSF Division of Astronomical Sciences. The award is made in recognition of significant contributions to Rubin Observatory’s Legacy Survey of Space and Time. 

\section{Author Contributions}

A.D.F. lead the \textit{TESS} data reduction, the creation of the deepstack images, the light curve extraction, manuscript writing, and figure creation. J.W.N. contributed manuscript text and review. D.Z.S. contributed manuscript text and review.

\appendix

In this appendix we include additional detail on our analysis. In particular, we focus on the background removal in Section~\ref{app:bkg} and a more in-depth discussion of our  aperture selections in Section~\ref{app:aperture}.
\vspace{3mm}

\section{Background Removal}\label{app:bkg}

\begin{figure}[ht!]
\includegraphics[width=1.0\linewidth, trim={0.1cm 0 0 0}, clip]{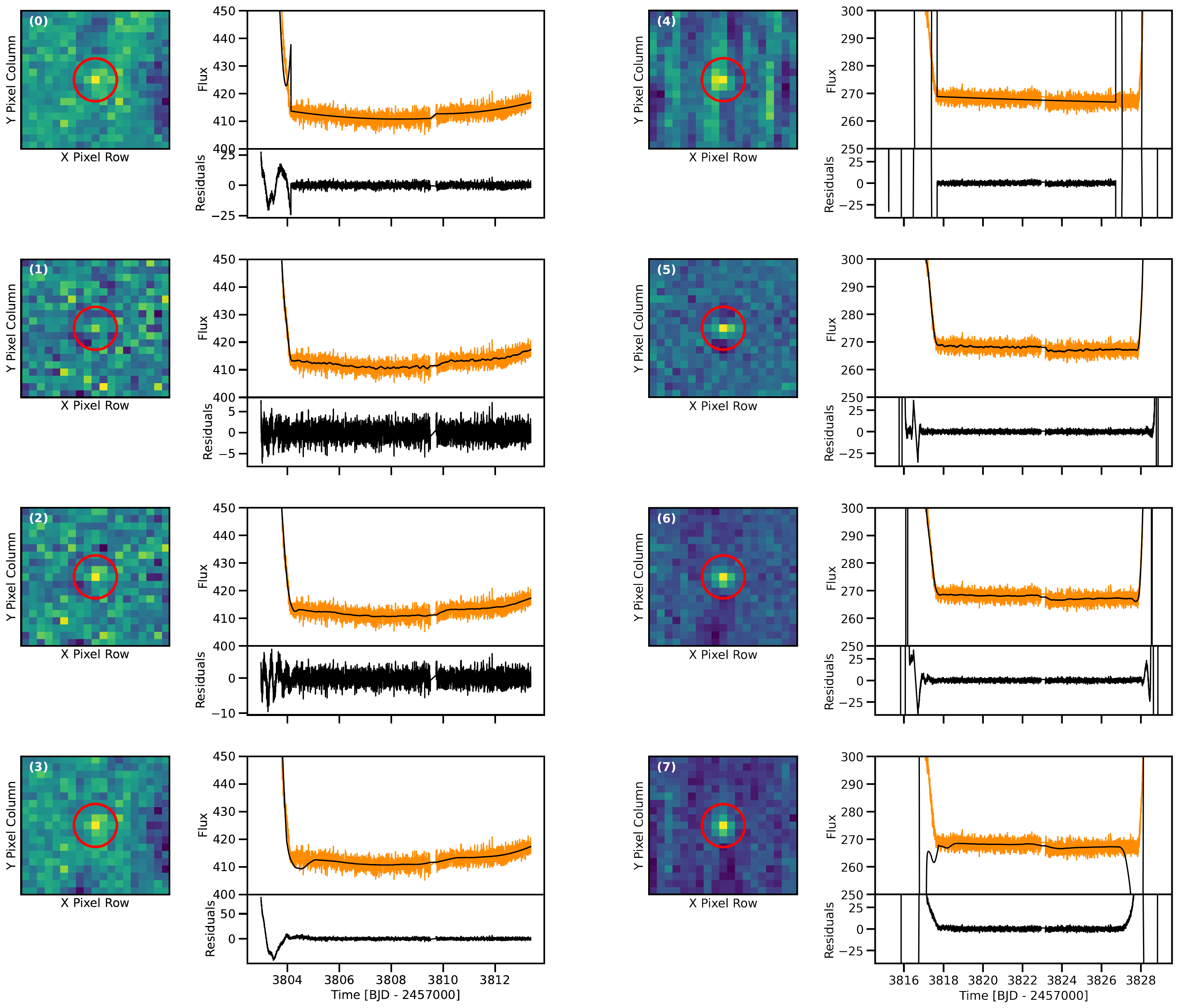}
\caption{An example of the various background fits to each pixel. Subplots (0-3) are representative of Camera 2 CCD 3; subplots (4-7) are representative of Camera 1 CCD 2. The extracted calibrated flux is shown in orange, the best-fit model overplotted in black, and the resulting residuals underneath in black. The left panel of each subpanel shows the median background-subtracted deepstack image for each model. The first row presents a 2\textsuperscript{nd}-order polynomial fit. The second, third, and fourth rows present a Savitsky-Golay filter model with window lengths of 121, 307, and 855, respectively.}
\label{fig:polynomial}
\end{figure}

In Figure~\ref{fig:polynomial} we present the models and residuals tested to remove the background properties in the \textit{TESS} FFIs. These experiments demonstrated that the Savitsky-Golay filter (with a window length of 307) presents the optimal fit to the observations \textit{without} overfitting the faint signal of 3I/ATLAS for Camera 2 CCD 3.  The larger window length is also able to recover the target but improperly fits the sharp decline of the orbital ramp. The 2\textsuperscript{nd}-order polynomial does not fit the Camera 1 CCD 2 data well due to the strong systematics present in the observations. Similarly to Camera 2 CCD 3, the second image is best fit by the Savitsky-Golay filter with a window length of 307. Therefore, we select this background to subtract from the calibrated FFIs in the entirety of the analysis presented in this work.

\section{Aperture Selection}\label{app:aperture}

The \textit{TESS} pixel response function (PRF) can change substantially over a given camera's FOV. Additionally, the PRF is slightly chromatic and can vary with temperature. The \textit{TESS} PRF model assumes that 24\% of the total source flux will be contained in the central pixel if the target is perfectly centered. If the target is offset, then the total amount of flux decreases. It is particularly important to take this into account when trying to extract the flux from a moving target. To combat these effects, larger apertures are more robust against these flux losses. However, implementing such large apertures is also particularly challenging for crowded fields because they may include flux contributions from nearby sources.

\begin{figure}[ht!]
\includegraphics[width=1.0\linewidth, trim={0.1cm 0 0 0}, clip]{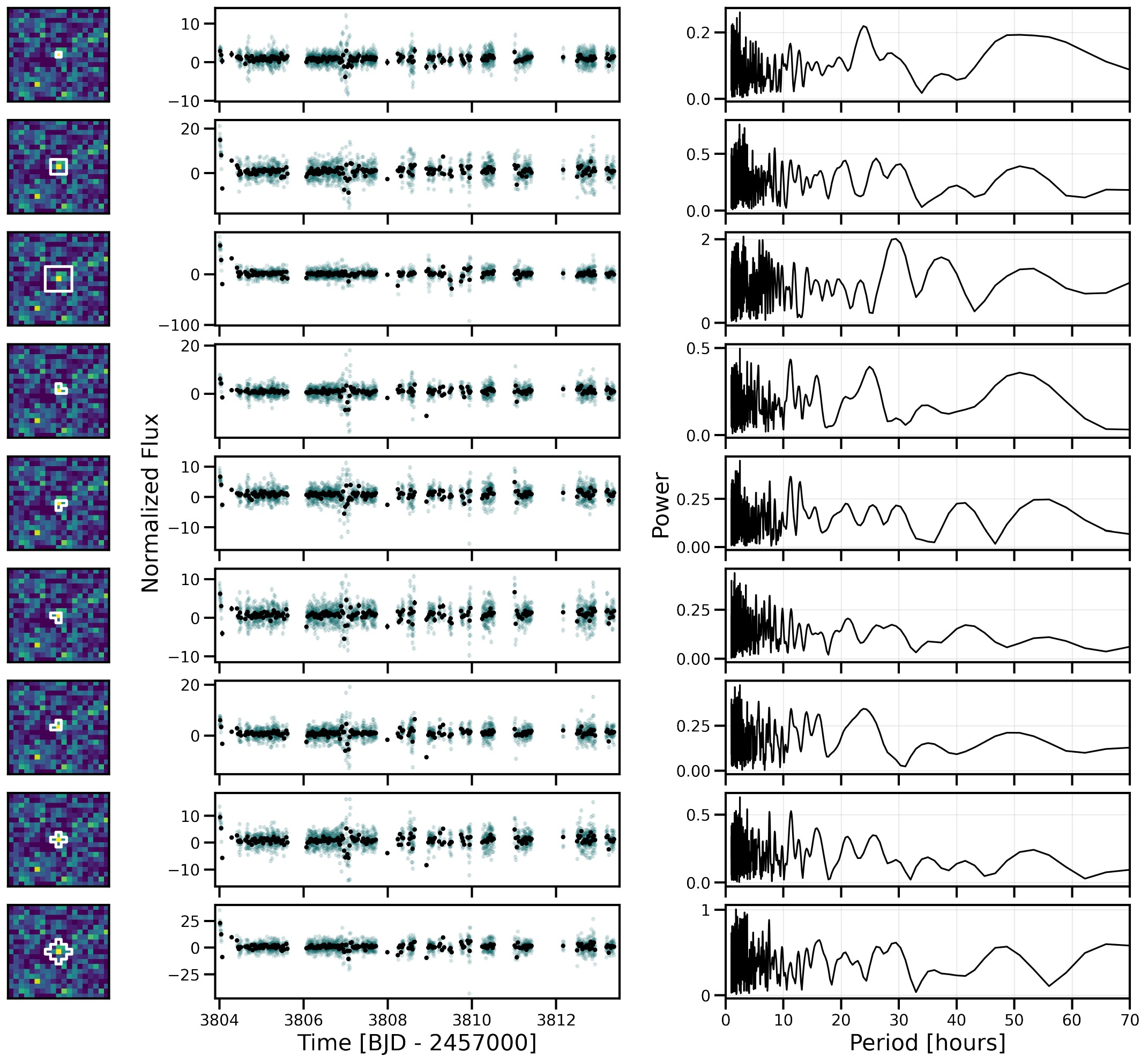}
\caption{Testing different apertures to create the light curve of 3I/ATLAS from Camera 2 CCD 3. The first column is the median background-subtracted deepstack image (same as Figure~\ref{fig:stacked}a) with a contour overlay of the aperture we tested. We choose this selection of apertures based on those presented in \cite{eleanor}. Each aperture includes the central pixel and at least two surrounding pixels. The image colormap scaled to Figure~\ref{fig:stacked}. The middle column is the extracted normalized light curve. The unbinned light curve is plotted in blue while the data binned to 36~minutes is plotted in black. The right column is the periodogram of the binned data.}
\label{fig:ap23}
\end{figure}

\begin{figure}[ht!]
\includegraphics[width=1.0\linewidth, trim={0.1cm 0 0 0}, clip]{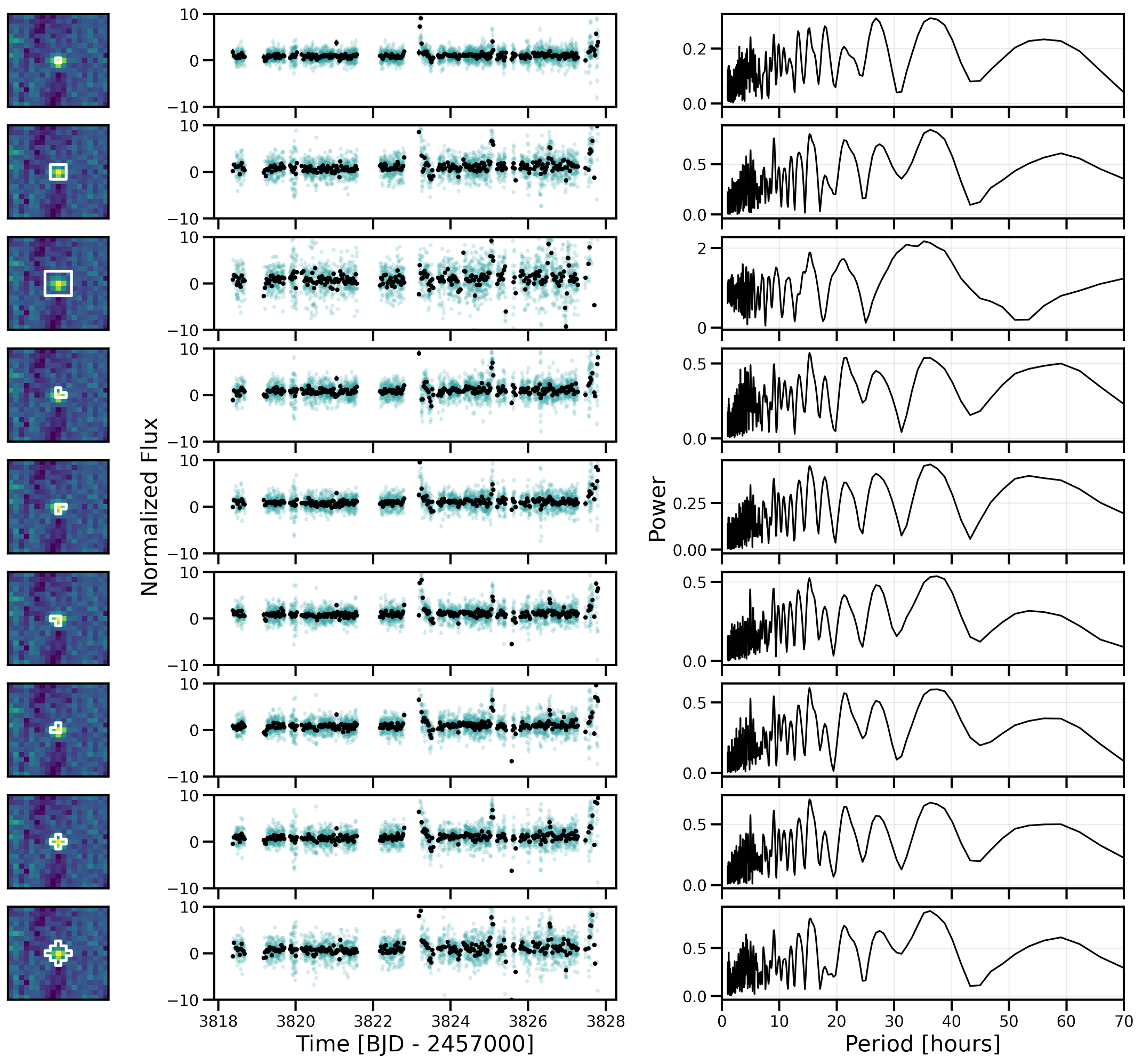}
\caption{Same as Figure~\ref{fig:ap23}, but for our second epoch of observations from Camera 1 CCD 2.}
\label{fig:ap12}
\end{figure}

To investigate the efficacy of both competing effects, we perform a set of experiments on a variety of  apertures to extract the light curve of 3I/ATLAS. The subset of apertures investigated here are provided in \cite{eleanor}, ranging  from a single pixel to a $3\times3$~area.  The results of these tests are presented in Figure~\ref{fig:ap23} and Figure~\ref{fig:ap12}. These figures show the full light curve as well as the light curve binned to 36~minutes. We also run the binned light curve through a Lomb-Scargle periodogram to search for potential evidence of a rotation period.

Figure~\ref{fig:ap23} demonstrates  that the light curve is dominated by noise at $P < 10$~hours. Additionally, there is no strong evidence of periodicity for $P > 30$~hours. While there are  stronger peaks at $P=15-28$~hours,  the majority of these periodograms have peaks at identical times within this time range. We note that the periodograms shown in rows 5-8 of Figure~\ref{fig:ap23} contain similar peaks at $\sim 19$~hours, which could be consistent with the reported $P_\textrm{rot} = 16.79 \pm 0.23$~hours \citep{Marcos2025}. However, because the same signal is not seen across all extracted light curves, we attribute it to instrumental systematics.

We repeat this same test with the data extracted from Camera 1 CCD 2. These results are presented in Figure~\ref{fig:ap12}. Similarly to the previous analysis of the other \textit{TESS} image,  the  $P<10$~hours is dominated by noise. Contrary to the previous analysis however, there is a strong peak between $37-41$~hours. Additionally, we find a consistent strong double-peaked feature at $P = 16, 22, 28$~hours. The 16~hour peak is consistent with $P_\textrm{rot}$ recovered by \citep{Marcos2025, santanaros25}. These signals are strong in every extracted light curve. None of the stronger peaks seen in the light curves extracted from Camera 1 CCD 2 are consistent with those from Camera 2 CCD 3. Additionally, because there are multiple strong peaks and no observable photometric variability in all data, it is challenging to determine whether the signals are instrumental or astrophysical in origin. Nevertheless, because the periodograms between \textit{TESS} data sets are inconsistent, we conclude that we are unable to measure a rotation period from these observations.

\bibliography{main}{}
\bibliographystyle{aasjournalv7}

\end{document}